\documentclass[twocolumn,reprint,epsf,graphics,psfig,superscriptaddress]{revtex4}

\usepackage{amsfonts}
\usepackage{graphicx}
\usepackage{amsmath}
\usepackage{amssymb}
\usepackage{latexsym}
\usepackage{psfrag}
\usepackage{dcolumn}
\usepackage{datetime}
\usepackage{multirow}
\usepackage{subfigure}
\usepackage{amssymb,amsmath,rotate,color}

\begin{document}
\title {Doping-controlled surface conduction in topological insulators with warping effects}
\author{Masomeh Arabikhah}
\affiliation{Department of Physics, Payame Noor University, P.O.Box 19395-3697 Tehran, Iran}
\author{Alireza Saffarzadeh}
\altaffiliation{Author to whom correspondence should be addressed. Electronic mail: asaffarz@sfu.ca}
\affiliation{Department of Physics, Payame Noor University, P.O.Box 19395-3697 Tehran, Iran} \affiliation{Department of Physics,
Simon Fraser University, Burnaby, British Columbia, Canada V5A
1S6}
\date{\today}

\begin{abstract}
Based on a self-consistent $t$-matrix approximation, we explore the influence of magnetic and nonmagnetic doping on the surface electronic states and conductivity of topological insulators. We show that warping parameter has a crucial impact on the density of states and dc conductivity of the doped surfaces.  As the warping strength is increased, the surface density of states at high energies is suppressed and the resonant states induced by impurities in the vicinity of the Dirac point gradually disappear. It is found that nonmagnetic impurities break electron-hole symmetry at low warping strength, while the symmetry remains unchanged when the surface is magnetically doped. Our findings reveal that surface conductivity can be controlled by tuning the doping, the direction of external magnetic field and that of impurity magnetic moments. Also, the surface conductivity features in topological insulators with warped energy dispersions are not significantly affected by the presence of impurities compared to that of materials with circular energy contour.

\end{abstract}
\maketitle
\section{INTRODUCTION}
Three-dimensional (3D) topological insulators are a special class of materials which possess a gapless spectrum at their surface as a consequence of time-reversal symmetry and band inversion induced by a strong spin-orbit coupling, while their bulk spectrum is fully gapped \cite{Kane2005,Hasan2010,Qi2011,Ando2013,Bansil2016}. The surface states of such systems contain only a single or odd number of Dirac cones
\cite{Berneving2006,Moore2007,Zhang2009}. The chiral nature of surface states has some peculiar implications on transport properties of these materials, which makes them an attractive candidate for quantum computation and spintronics devices \cite{Murakami2004,Fu2008,Tse2010,Culcer2012,Aguilar2015,wang2015}.

The surface states are protected against local perturbations that preserve time-reversal symmetry \cite{Fu2007}. One way to examine the topological stability is to probe the energy spectrum of topological insulators in the presence of randomly distributed ionized and magnetized impurities. The conducting surface states of these materials are robust against the effect of non-magnetic disorder, while the Dirac surface state spectrum can be gapped by magnetic impurities or other perturbations breaking time-reversal symmetry \cite{Liu2009,Biswas2010,Black2012,Black2015,Wozny2018,pieper2016,Zhong-Li-2017,Sabzalipour2015}. Nevertheless, a well-defined band-gap is not always formed by magnetic impurities, as observed in some experiments \cite{Sessi2016,Krieger2017,Yu2020}.

The continuum surface model and also a three-dimensional tight-binding lattice model, showed that the energy gap induced by ferromagnetically
aligned magnetic impurities on the surface of a topological insulator can be filled, due to the scattering off the nonmagnetic potential of the impurities that induce low-energy impurity resonances \cite{Black2015}. The influence of random impurities on helical edge states for a fixed magnetic contribution has been studied \cite{Wozny2018}. It was shown that the gap is filled when either the scalar potential or Fermi velocity is increased \cite{Wozny2018}.
Based on kernel polynomial expansion and Green's function approaches, the local density of states, single-particle spectral function, and electric  conductance on the surface of 3D topological insulators have been investigated  \cite{pieper2016}. It was shown that bulk disorder refills the surface-states band gap induced by a homogeneous magnetic field, while random Peierls phases preserve the gap feature \cite{pieper2016}.
Zhong et al., \cite{Zhong-Li-2017}  studied the optical and thermoelectric properties of a topological insulator surface in the presence of impurities and also the role of resonant scattering within the full Born approximation. It was demonstrated that although the scalar impurities cannot lead to a perfect backscattering, they can induce particle-hole asymmetry and resonant states through skew scattering \cite{Zhong-Li-2017}. Also, using the semiclassical Boltzmann approach and a generalized relaxation time approximation it was shown that the surface conductivity is anisotropic and strongly dependent on both the direction of impurity spin and the magnitude of bulk magnetization \cite{Sabzalipour2015}.

The electronic states of low energy electrons in topological insulators with circular Fermi surface can be well described by Dirac equation.
However, by increasing the Fermi energy in some materials, the constant-energy contour (CEC) is distorted from circular to hexagonal and in some cases to snowflake. In this regard, the angle-resolved photoemission spectroscopy experiments and the band structure calculations have shown that the Fermi surface of Bi$_{2}$Te$_{3}$ has a single Dirac cone on its surface and is distorted into hexagonal snowflake as the Fermi energy is increased \cite{Chen2009,Li-Carbotte-2013,leBlance2014}. This phenomenon that is known as hexagonal warping effect introduces nonlinear corrections to energy dispersion away from the Dirac point and describes cubic spin-orbit coupling at the surface of rhombohedral crystal systems \cite{Fu2009,Nomura2014}. Therefore, not all the surface states of topological insulators exhibit the hexagonal warping effects. The size of energy band gap and the crystal symmetry of topological insulators are crucial in their hexagonal warping strength of surface band dispersions.

The energy dependence of electronic density of states (DOS) in the presence of warping effect deviates from linearity with increasing the electron energy \cite{Li-Carbotte-2013}. The electrical conductivity of topological insulators, quantum anomalous Hall effect, anomalous out-of-plane magnetoresistance \cite{Akzyanov2018}, and also the real part of longitudinal conductivity with non-magnetic impurities are all affected by the warping strength \cite{chen2018}. Moreover, the hexagonal warping gives rise to additional anisotropic components in the spin conductivity \cite{Akzyanov2019}.

In this paper, we investigate the electronic DOS and charge conductivity on the surface of magnetized topological insulators with warped energy bands in the presence of magnetic and nonmagnetic impurities. Using a self-consistent $t$-matrix approximation, we show that a magnetically induced gap can be filled by nonmagnetic doping. The electric dc conductivity depends on the hexagonal warping parameter, impurity concentration and exchange field strength. Moreover, the conductivity and time-reversal symmetry on the surface of doped topological insulators are strongly affected by changing the magnetization orientation. The broken time-reversal symmetry, induced by an external magnetic field and/or impurity magnetic moments, manifests itself as a gap opening in energy dispersion and electric conductivity. 

This paper is organized as follows. In Se. II, we present our model and formalism for the surface states of a doped topological insulator with warped energy bands in the presence of magnetic proximity effect. Our numerical results in the self-consistent $t$-matrix approximation for the surface DOS and charge conductivity are discussed in Sec. III. Finally, the conclusion is given in Sec. IV.

\section{MODEL AND FORMALISM}
We consider a 3D topological insulator such as $\mathrm{Bi}_{2}\mathrm{Te}_{3}$ with strong warping effects and a Dirac cone on its surface. The interaction between bulk states and the surface states can be ignored by tuning the position of Fermi level on the surface states by means of an external gate voltage \cite{Arabikhah2019}  or an appropriate doping \cite{Chen2009}. The effective Hamiltonian of the surface states in the presence of hexagonal warping and the absence of particle-hole asymmetry can be given in units of $\hbar=1$ as \cite{Fu2009,Chiba-Takahashi-2017}
\begin{equation}\label{eq1}
\hat{H}=\upsilon_{F}(k_{x}\sigma_{y}-k_{y}\sigma_{x})+\lambda(k_{x}^{3}-3k_{x}k_{y}^{2})\sigma_{z}+\Delta\mathrm{\bf{M}}\cdot\boldsymbol{\sigma} \ ,
\end{equation}
where the first term describes the helical Dirac fermions on the surface of the topological insulator, the second term represents the hexagonal warping effect (see Fig. 1(a)), and the third term denotes the effect of magnetic proximity. Also, $k_x=k\cos\beta$ and $k_y=k\sin\beta$ are the in-plane momentum components, $\upsilon_{F}$ is the Fermi velocity, $\lambda$ is the warping parameter, and $\boldsymbol{\sigma}=(\sigma_{x},\sigma_{y},\sigma_{z})$ are the Pauli matrices. 

To induce the magnetic proximity effect, the surface of 3D topological insulator is coated by a ferromagnetic insulator (FMI) layer, such as EuS, with magnetization direction ${\bf M}=(\sin\theta\cos\varphi,\sin\theta\sin\varphi,\cos\theta)$, where $\theta$ is the polar angle with respect to the $z$-axis and $\varphi$ is the azimuthal angle measured from the $x$-axis, as shown in Fig. 1(b). The proximity-induced magnetization at the surface of topological insulator couples to the spin of surface electron with the coupling strength $\Delta$. Therefore, the short-range nature of this magnetic coupling allows the surface states to experience a ferromagnetic interaction which influences the energy spectrum of the topological states, rather than affecting the bulk states \cite{Wei2013}.

The energy dispersion of the topological surface states can be obtained by
\begin{equation}\label{eq2}
E({\bf k})=s\sqrt{\upsilon_{F}^2k^{2}+\Delta^{2}\sin^{2}{\theta}-A+B}\, ,
\end{equation}
where $A=2\upsilon_{F}k\Delta\sin\theta\sin(\beta-\varphi)$,
$B=({\lambda}k^{3}\cos3\beta+\Delta\cos\theta)^2$, and $s={\pm 1}$ correspond to the upper and lower surface bands, respectively (see Fig. 1(a)).

The magnetic and chemical (nonmagnetic) impurities, acting as point-like potentials randomly distributed at the positions ${\bf r}_{n}$ on the surface of topological insulator (see Fig. 1(b)) can be expressed in the framework of classic theory as \cite{Zhong-Li-2017,Saffar-2008}
\begin{equation}\label{eq3}
V_{imp}(\mathbf{r})=\sum_{n}(U\boldsymbol{\sigma_{0}}+J\mathrm{\bf{m}}\cdot\boldsymbol{\sigma}) \delta_{\mathbf{r},\mathbf{r_{n}}}\, ,
\end{equation}
where $U$ is a scalar potential with the identity matrix
$\boldsymbol{\sigma_{0}}$, while $J=I S$ is the exchange coupling strength between Fermi electrons and the spin, $S$, of local magnetic
impurities with magnetic direction ${\bf{m}}=(\sin\theta_{m}\cos\varphi_{m},\sin\theta_{m}\sin\varphi_{m},\cos\theta_{m})$. Here, $\theta_{m}$ is the polar angle measured from the $z$-axis, $\varphi_{m}$ is the azimuthal angle with respect to the $x$-axis and $I$ is the exchange constant. The magnetic impurities in the presence of proximity effect are assumed to be aligned in the same direction of external magnetic field, such that $\mathrm{\bf{m}}=\mathrm{\bf{M}}$.
Therefore, the total Hamiltonian is given by $\hat{\mathcal{H}}=\hat{H}+V_{imp}(\mathbf{r})$, where we have added the effects
of magnetic and nonmagnetic impurities to the unperturbed Hamiltonian $\hat{H}$.

To consider the effects of impurity scattering, we calculate the total Green's function, $\mathcal{G}(\omega,\mathbf{k})$, using the Dyson equation
 \begin{equation}\label{eq4}
\mathcal{G}^{-1}(\omega,\mathbf{k})=G^{-1}(\omega,\mathbf{k})-\Sigma(\omega)\, .
\end{equation}
Here, $\omega$ is the electron energy, $\Sigma(\omega)$ is the self-energy which includes the effect of all scattering processes, and $G(\omega)=(\omega-\hat{H})^{-1}$ is the unperturbed Green$^{,}$s function that can be written as a $2 \times 2$ matrix:
\begin{equation}\label{eq5}
G(\omega,\mathbf{k})=
\begin{pmatrix}
G_{11}(\omega,\mathbf{k}) & G_{12}(\omega,\mathbf{k})\\
G_{21}(\omega,\mathbf{k}) & G_{22}(\omega,\mathbf{k})
\end{pmatrix},
\end{equation}
where the diagonal and off-diagonal elements are given by
\begin{equation}\label{eq6}
G_{11(22)}(\omega,\mathbf{k})=\frac{\omega\pm\lambda(k_{x}^{3}-3k_{x}k_{y}^{2})\pm\Delta\cos\theta}{D}\, ,
\end{equation}
 \begin{equation}\label{eq7}
G_{12(21)}(\omega,\mathbf{k})=\frac{\upsilon_{F}(\mp i k_{x}-k_{y})+e^{\mp i \varphi} \Delta\sin\theta}{D}\, ,
\end{equation}
with $D=\omega^{2}-(\upsilon_{F}k_{x}+\Delta\sin\theta\sin\varphi)^{2}-(\upsilon_{F}k_{y}-\Delta\sin\theta\cos\varphi)^{2}\nonumber-(\lambda(k_{x}^{3}-3k_{x}k_{y}^{2})+\Delta\cos\theta)^{2}$. Here, the upper signs correspond to the elements $G_{11}(\omega,\mathbf{k})$ and $G_{12}(\omega,\mathbf{k})$, while the lower signs correspond to $G_{21}(\omega,\mathbf{k})$ and $G_{22}(\omega,\mathbf{k})$.
\begin{figure}
\centerline{\includegraphics[scale=0.3]{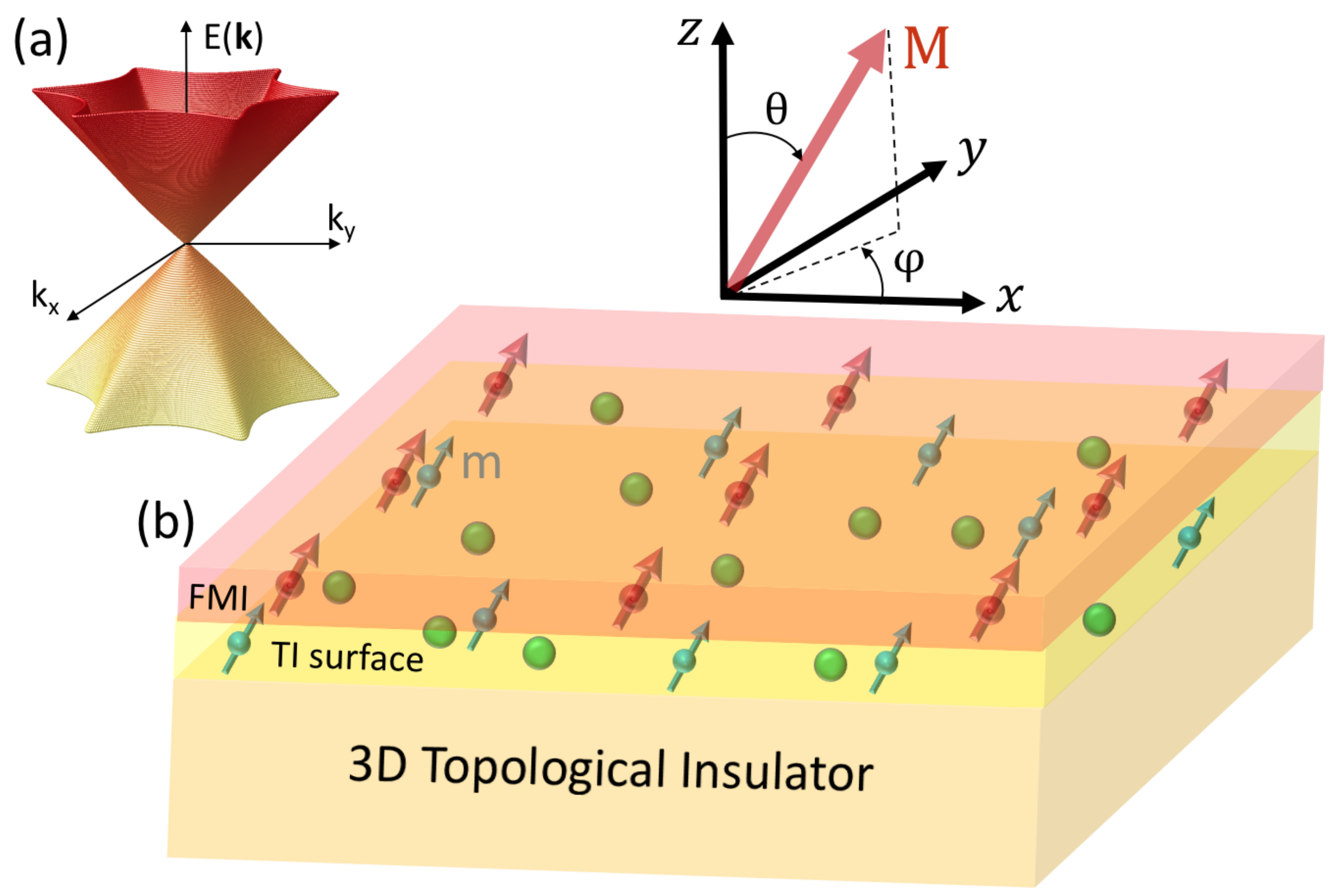}}
\caption{(a) The Dirac cone of fermions with Fermi energy $E=0.4$ eV on the surface of a 3D topological insulator (TI) in the presence of hexagonal warping effect. (b) Schematic view of a 3D topological insulator with random distribution of magnetic impurities (blue spheres) with magnetic direction $\bf m$ and nonmagnetic impurities (green spheres) on the TI surface in the presence of magnetic proximity effect induced by the FMI layer with magnetization direction $\bf M$. The regular array of brown spheres (arrows) in the FMI layer represents the magnetic atoms (moments). }
\label{F1}
\end{figure}

There are various approximation schemes for the self-energy calculations. In the Born approximation, which is called here $t$-matrix approximation, the electrons are only scattered off one impurity. In the self-consistent Born approximation, also known as self-consistent $t$-matrix approximation, an infinite number of impurities appears, thereby improving the approximation. It will allow us to study the limits of weak and strong scatterers. The difference  between $t$-matrix approximation and self-consistent $t$-matrix approximation is that the former uses the unperturbed Green's function, while in the latter the impurity averaged Green's function is used in the $t$-matrix expansion. In the self-consistent $t$-matrix approximation, the $t$-matrix expansion is given by \cite{Wozny2018,Ostrovsky2006,Peres2006}
\begin{equation}\label{eq8}
t(\omega)=V_{imp}+V_{imp} \sum_{\mathbf{k}}\mathcal{G}(\omega,\mathbf{k}) t(\omega).
\end{equation}
Moreover, the self-energy which is defined as $\Sigma(\omega)=n_{imp} t(\omega)$, can be rewritten as
\begin{equation}\label{eq10}
\Sigma(\omega)=n_{imp}(\sigma_{0}-V_{imp} g(\omega))^{-1} V_{imp}\, ,
\end{equation}
where, $n_{imp}=N_{imp}/N$ is the concentration of impurities with $N_{imp}(N)$ being the total number of impurity (host) atoms, $V_{imp}=U\boldsymbol{\sigma_{0}}+J\mathrm{\bf{m}}\cdot\boldsymbol{\sigma}$ and $g(\omega)$ being the integral of the Green's function,
\begin{equation}\label{eq11}
g(\omega)=\int \frac{d^{2}k}{(2\pi)^{2}}\mathcal{G}(\omega,\mathbf{k}) .
\end{equation}
The self-energy matrix $\Sigma(\omega)$, consisting of diagonal elements ($\Sigma_{11}(\omega), \Sigma_{22}(\omega)$) and off-diagonal elements ($\Sigma_{12}(\omega), \Sigma_{21}(\omega)$) is determined self-consistently by Eqs. (\ref{eq4}), (\ref{eq10}), and (\ref{eq11}). Then, the total Green's function can be computed as
\begin{equation}\label{eq12}
\mathcal{G}(\omega,\mathbf{k})=
\begin{pmatrix}
\mathcal{G}_{1}(\omega,\mathbf{k})& \mathcal{G}_{2}(\omega,\mathbf{k})\\
\mathcal{G}_{3}(\omega,\mathbf{k})& \mathcal{G}_{4}(\omega,\mathbf{k})
\end{pmatrix},
\end{equation}
where the diagonal terms are given by
\begin{equation}\label{eq13}
\mathcal{G}_{1(4)}(\omega,\mathbf{k})=\frac{G_{11(22)}(\omega,\mathbf{k})+h\Sigma_{22(11)}(\omega)}{1-(f+hl)}\, ,
\end{equation}
and the off-diagonal terms are
\begin{equation}\label{eq14}
\mathcal{G}_{2(3)}(\omega,\mathbf{k})=\frac{G_{12(21)}(\omega,\mathbf{k})-h\Sigma_{12(21)}(\omega)}{1-(f+hl)}\, ,
\end{equation}
with $f=\Sigma_{11}(\omega)G_{11}(\omega,\mathbf{k})+\Sigma_{22}(\omega)G_{22}(\omega,\mathbf{k})+
\Sigma_{12}(\omega)G_{21}(\omega,\mathbf{k})+\Sigma_{21}(\omega)G_{12}(\omega,\mathbf{k})$ , $l=\Sigma_{11}(\omega)\Sigma_{22}(\omega)-\Sigma_{21}(\omega)\Sigma_{21}(\omega)$ and $h=G_{12}(\omega,\mathbf{k})G_{21}(\omega,\mathbf{k})-G_{11}(\omega,\mathbf{k})G_{22}(\omega,\mathbf{k})$.

Calculating the self energy and the Green's function, we can find the DOS and conductivity of the system. The density of state can be obtained by
\begin{equation}\label{eq15}
\rho(\omega)=\mp\frac{1}{\pi}\mathrm{Im}\,\mathrm{Tr}\int\frac{d^{2}k}{(2\pi)^{2}}\mathcal{G}^{\pm}(\omega,\mathbf{k})\, ,
\end{equation}
where $\mathcal{G}^{\pm}(\omega)=(\omega\pm i\delta-\hat{H}-\Sigma)^{-1}$ are the retarded and advanced total Green's function, respectively, in which $\delta$ is a positive infinitesimal. Furthermore, a fully quantum mechanical expression for the conductivity in linear response regime is provided by the Kubo theory. In this regard, the longitudinal conductivity at zero temperature can be presented as a sum of the three terms $\sigma_{xx}=\sigma_{xx}^{\mathcal{\mathrm{I}}}+\sigma_{xx}^{\mathrm{II}}+\sigma_{xx}^{\mathrm{III}}$ where
\begin{equation}\label{eq16}
\sigma_{xx}^{\mathrm{I}}=\frac{e^{2}}{2\pi}\\\mathrm{Tr} \langle v_{x}\mathcal{G}^{+} v_{x}\mathcal{G}^{-}\rangle_{c}\, ,
\end{equation}
\begin{equation}\label{eq17}
\sigma_{xx}^{\mathrm{II}}=-\frac{e^{2}}{4\pi}\\\mathrm{Tr} \langle v_{x}\mathcal{G}^{+}v_{x}\mathcal{G}^{+}+v_{x}\mathcal{G}^{-}v_{x}\mathcal{G}^{-}\rangle_{c}\, ,
\end{equation}
\begin{eqnarray}\label{eq18}
\sigma_{xx}^{\mathrm{III}}&=&\frac{e^{2}}{4\pi}\int_{-\infty}^{+\infty}d\omega f(\omega)\mathrm{Tr}[v_{x}\mathcal{G}^{+} v_{x}\frac{d\mathcal{G}^{+}}{d\omega}-v_{x}\frac{d\mathcal{G}^{+}}{d\omega}v_{x}\mathcal{G}^{+}\nonumber\\
&&-v_{x}\mathcal{G}^{-} v_{x}\frac{d\mathcal{G}^{-}}{d\omega}+v_{x}\frac{d\mathcal{G}^{-}}{d\omega}v_{x}\mathcal{G}^{-}]\ ,
\end{eqnarray}
Here, $v_{x}=\frac{\partial\hat{H}}{\partial k_{x}}$ is the velocity operator, $f(\omega)$ is the Fermi distribution, and the subscript $c$ indicates the average over all disordered configurations. $\sigma_{xx}^{\mathrm{I}}$, $\sigma_{xx}^{\mathrm{II}}$ correspond to the contribution of states at the Fermi surface, while $\sigma_{xx}^{\mathrm{III}}$ contains the contribution of all filled states below the Fermi energy.
Since we study the conductivity of electrons at the Fermi surface, we only need to consider Eqs. (\ref{eq16}) and (\ref{eq17}).
Therefore, the conductivity of the topological insulator is given by the Kubo-Bastin-Streda formula \cite{Proskurin2015}
\begin{equation}\label{eq19}
\sigma_{xx}=\frac{e^{2}}{\pi} \mathrm{Tr} \int\frac{d^{2}k}{(2\pi)^{2}}[v_{x}\mathrm{Im}\mathcal{G}(\omega,\mathbf{k})v_{x}\mathrm{Im}\mathcal{G}(\omega,\mathbf{k})]\, ,
\end{equation}
where $\mathrm{Im}\mathcal{G}(\omega,\mathbf{k})=\frac{i}{2}[\mathcal{G}^{+}(\omega,\mathbf{k})-\mathcal{G}^{-}(\omega,\mathbf{k})]$.
It should be mentioned that in the case of short-ranged scatterers, the current vertex correction vanishes \cite{Shon1998,Bastin1971,Ostrovsky2006,Zhong-Li-2017,Peng}. Moreover, in the presence of magnetic proximity, the difference between longitudinal conductivity with and without vertex correction is reduced \cite{ Chiba-Takahashi-2017}. Therefore, we have neglected the effect of vertex corrections in the conductivity calculations.

\begin{figure}
\centerline{\includegraphics[scale=0.44]{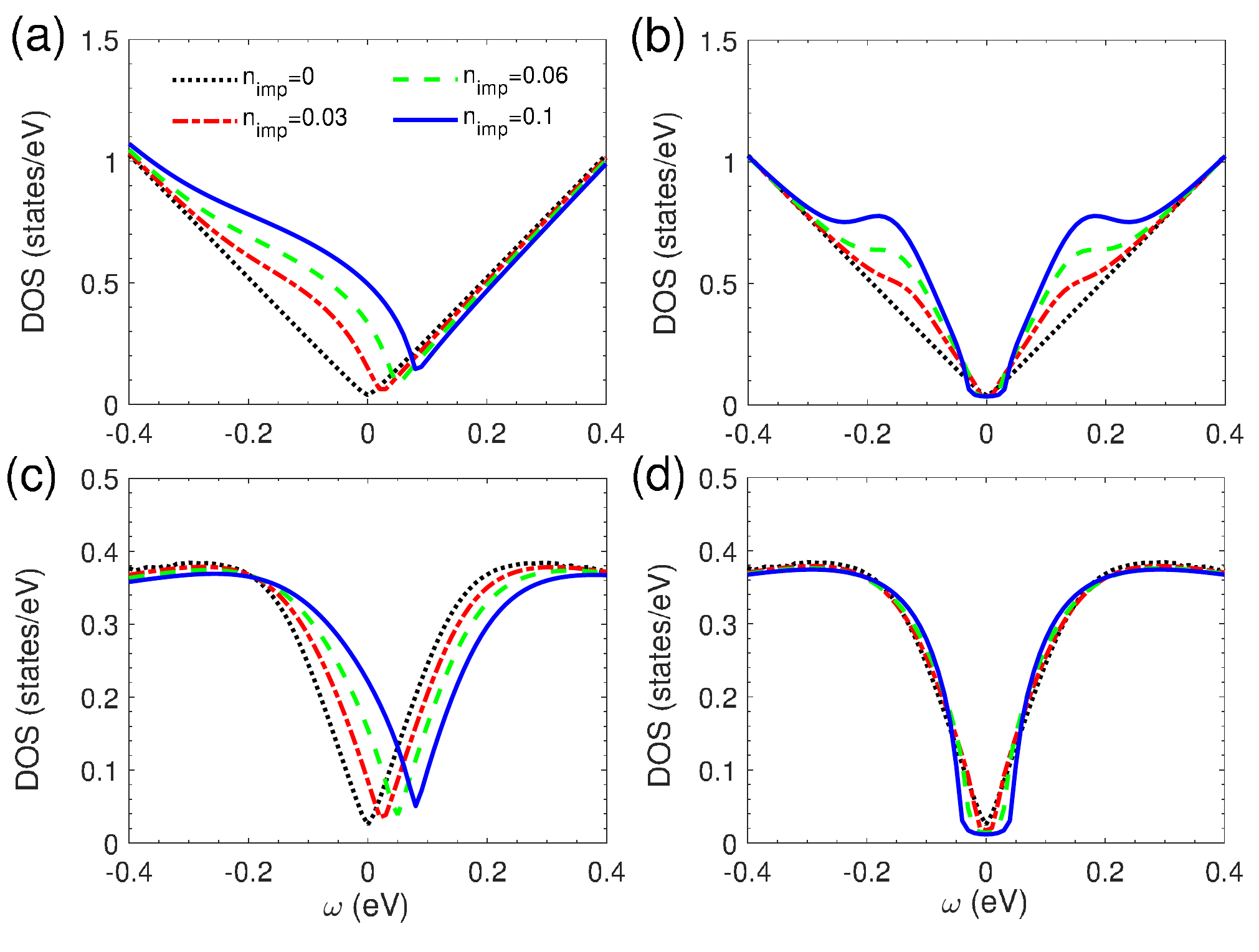}}
\caption{The surface DOS as a function of energy $\omega$ for different values of  $n_{imp}$ in the [(a), (b)] absence and [(c), (d)] presence of warping effect, when $\Delta=0$. The other parameters are (a) $\lambda=0$, $U=0.8$\,eV, $J=0$, (b) $\lambda=0$, $U=0$, $J=0.8$\,eV, (c) $\lambda=250$ eV$\cdot$\AA$^3$, $U=0.8$\,eV, $J=0$, and (d) $\lambda=250$\,eV$\cdot$\AA$^3$, $U=0$, $J=0.8$\,eV.}
\label{F2}
\end{figure}

\begin{figure}
\centerline{\includegraphics[scale=0.44]{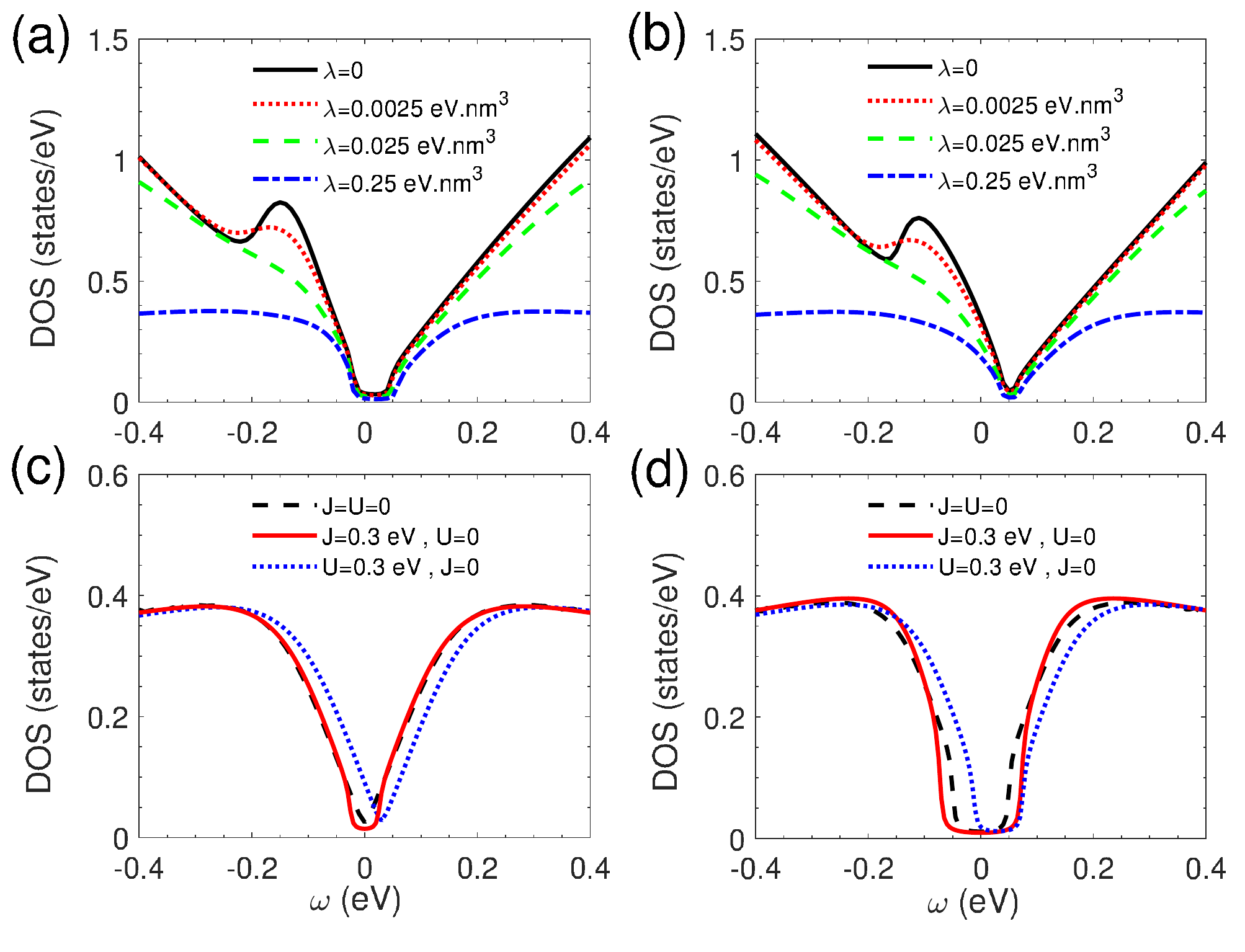}}
\caption{The surface DOS as a function of energy $\omega$ at impurity concentration $n_{imp}=0.1$. The other parameters are (a) $\Delta=0$, $J=0.6$ eV, $U=0.3$ eV, (b) $\Delta=0$, $J=0.3$ eV, $U=0.6$ eV, (c) $\Delta=0$, $\lambda=250$ eV$\cdot$\AA$^3$, and (d) $\Delta=0.05$ eV, $\lambda=250$ eV$\cdot$\AA$^3$.}
\label{F3}
\end{figure}

\section{RESULTS AND DISCUSSION}
Based on the above formalism, we present the numerical results of electronic states and longitudinal conductivity on the surface of doped topological insulators. We have taken $\upsilon_{F}=2.55$\,eV$\cdot$\AA\,and $\lambda=250$\,eV$\cdot$\AA$^3$ which produce the Fermi surface of Bi$_{2}$Te$_{3}$ in good agreement with experiment \cite{Chen2009}. In Figs. 2(a)-(b) and 2(c)-(d), we have depicted the surface electronic DOS for different values of surface impurity concentration $n_{imp}$ as a function of energy $\omega$ at $\lambda=0$ and $\lambda=250$ eV$\cdot$\AA$^3$, respectively. In the case of $\lambda =0$, the DOS shows a V-shaped behavior in the given energy window, reflecting a linear dispersion in the absence of any impurities. The introduction of nonmagnetic impurities with positive $U$ values ($p$-doping) on the surface, however, shifts the Dirac point to the higher-energy side, leading to an electron-hole asymmetry accompanied by a broaden resonance  at the energies lower than the Dirac energy (see Fig 2(a)). The increase in the impurity concentration $n_{imp}$, enhances the electronic states at the Dirac point, providing more carriers contributing in charge transport. For negative potential $U$ values ($n$-doping), the same resonant states appear but at the positive side of energy (not shown), in agreement with scanning tunneling microscopy experiment. \cite{Xu-2017}. In contrast, the magnetic impurities with magnetic direction  $\bf m$ along $z$-direction induce resonant states at both side of the Dirac point energy, preserving the electron-hole symmetry and leading to a band gap around the Dirac point whose size is dependent on $n_{imp}$ value, as shown in Fig. 2(b). Moreover, an increase in the impurity concentration can lead to an increase in the peak of resonant states, indicating that the electronic states in the vicinity of Dirac point can be tuned by varying the density of magnetic impurities. Compared to Figs. 2(a) and 2(b), the introduction of warping term in the Hamiltonian greatly modifies the surface states not only in a clean surface but also in doped surfaces with nonmagnetic and magnetic impurities, as shown in Figs. 2(c) and 2(d), respectively. The warping effect becomes more pronounced at energies above $|\omega|\simeq 0.2$ eV, where the surface DOS is suppressed dramatically. In this case, although the nonmagnetic impurities still shift the Dirac point, the resonant states induced by impurities almost disappear, preserving the electron-hole symmetry, regardless of the dopant type (see Figs. 2(c) and 2(d)). However, the gap opening around the Dirac point energy in the magnetically doped surfaces is not considerably affected by introduction of warping effect (see Figs. 2(b) and 2(d)).
Since the warping term depends on $\sigma_{z}$ and that $\bf m$ is along $z$ direction, the gap sizes are larger in Fig. 2(d) compared to those in Fig. 2(b).
\begin{figure}
\centerline{\includegraphics[scale=0.6]{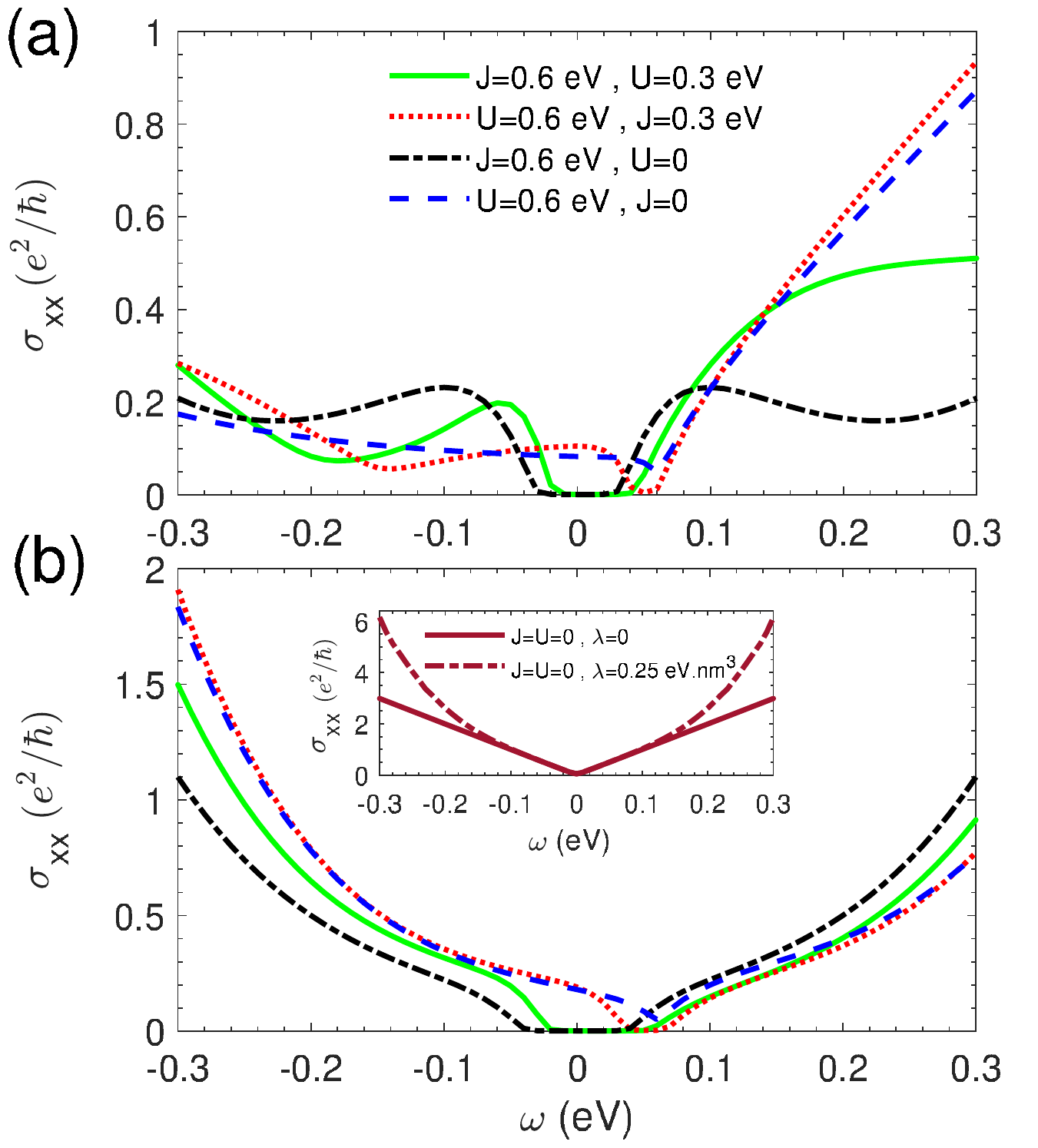}}
\caption{Calculated dc conductivities as a function of energy $\omega$ in the (a) absence and (b) presence of warping effect with strength $\lambda=250$ eV$\cdot$\AA$^{3}$. The impurity concentration is $n_{imp}=0.1$ and $\Delta=0$. The inset in (b) shows the conductivity of clean surfaces without (solid line) and with (dashed line) warping effect.}
\label{F4}
\end{figure}
\begin{figure}
\centerline{\includegraphics[scale=0.6]{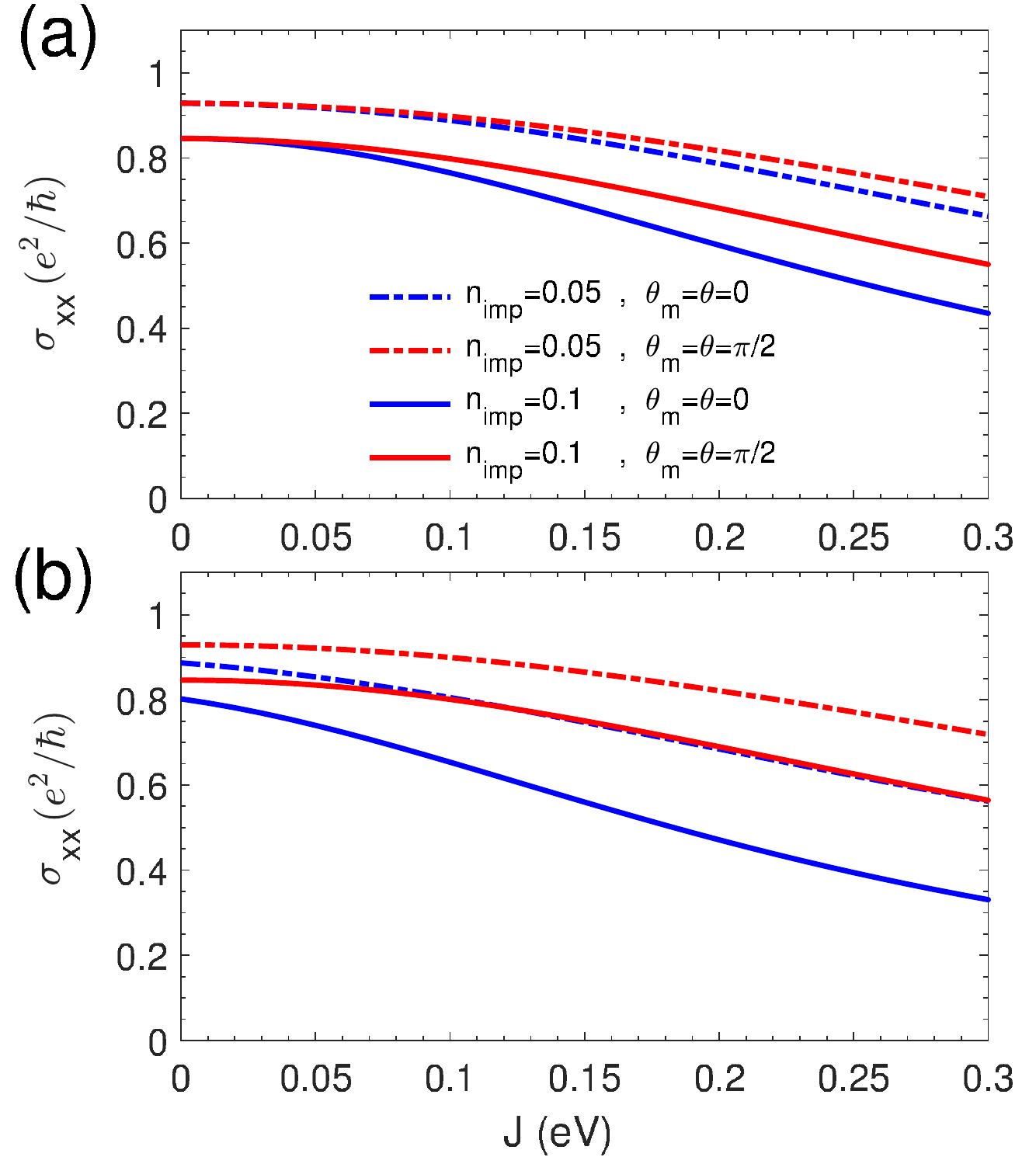}}
\caption{Calculated dc conductivities as a function of magnetic potential $J$ at (a) $\Delta=0$ and (b) $\Delta=0.02$\, eV. The other parameters are $U=0.1$\, eV, $\omega=0.1$\, eV, $\lambda=250$ eV$\cdot$\AA$^3$, and $\varphi_{m}=\varphi=0$.}
\label{F5}
\end{figure}

We have shown in Figs. 3(a) and 3(b) the surface DOS versus energy for different values of warping strength $\lambda$, when both magnetic and nonmagnetic impurities are present. The increase in $\lambda$ suppresses the electronic states gradually, and hence,
the particle-hole symmetry is restored. These features from the impurity effect are independent of the dopant type, and that whether $J>U$ or $U>J$ (compare Fig. 3(a) and 3(b)). Therefore, although the increase in the exchange coupling strength $J$ results a larger band gap opening and that the increase in the potential strength $U$ causes a larger shift in energy at the Dirac point, the particle-hole symmetry at low energies is always preserved in topological insulators with strong enough warping effect, such as Bi$_{2}$Te$_{3}$. In Figs. 3(c) and 3(d), we have depicted the surface DOS with warping parameter $\lambda=250$ eV$\cdot$\AA$^3$, in the absence and presence of magnetic proximity effect $\Delta=0.05$ eV in $z$ direction, respectively. When the surface is clean and $\Delta=0$, the electronic states represent a V-shaped behavior in the range of energy $|\omega|\simeq0.1$ eV, whereas the proximity effect opens up an energy gap of $2\Delta$ in DOS at the Dirac point (compare black dashed line in Figs. (3c) and 3(d)). The introduction of nonmagnetic impurities shifts the Dirac point in energy without any increase in the electronic states or any change in the energy gap. The introduction of magnetic impurities, however, induces an energy gap in the absence of proximity effect and increases the gap size in the presence of this effect. This feature can be explained by comparing the proximity term in Eq. (2) and the self-energy term in Eq. (9). When $\Delta=0$ and $J\neq 0$, the magnetic impurities still affect the Dirac fermions via self-energy term and create broken time-reversal symmetry, giving rise to the gap opening at the Dirac point. On the other hand, when $\Delta\neq 0$ and $J\neq 0$, not only the magnetic impurities, but also the proximity term in Eq. (2) affect collectively the Dirac fermions, causing a larger energy gap. Note that the magnetic proximity term and the self-energy term describe two separate phenomena. As mentioned above, the magnetic proximity effect induces a magnetization at the surface of topological insulator that couples to the spin of Dirac electrons, while the self-energy term contains the exchange coupling between localized magnetic moments and the spin of Dirac fermions. Therefore, the time-reversal symmetry is broken by both the proximity effect and magnetic impurities, whether or not the energy band is warped. Also, we see that as a result of warped energy band, the suppression of electronic states and the vanishing of resonant peak around the Dirac point are still retained, when an external magnetic field is applied.

\begin{figure}
\centerline{\includegraphics[scale=0.6]{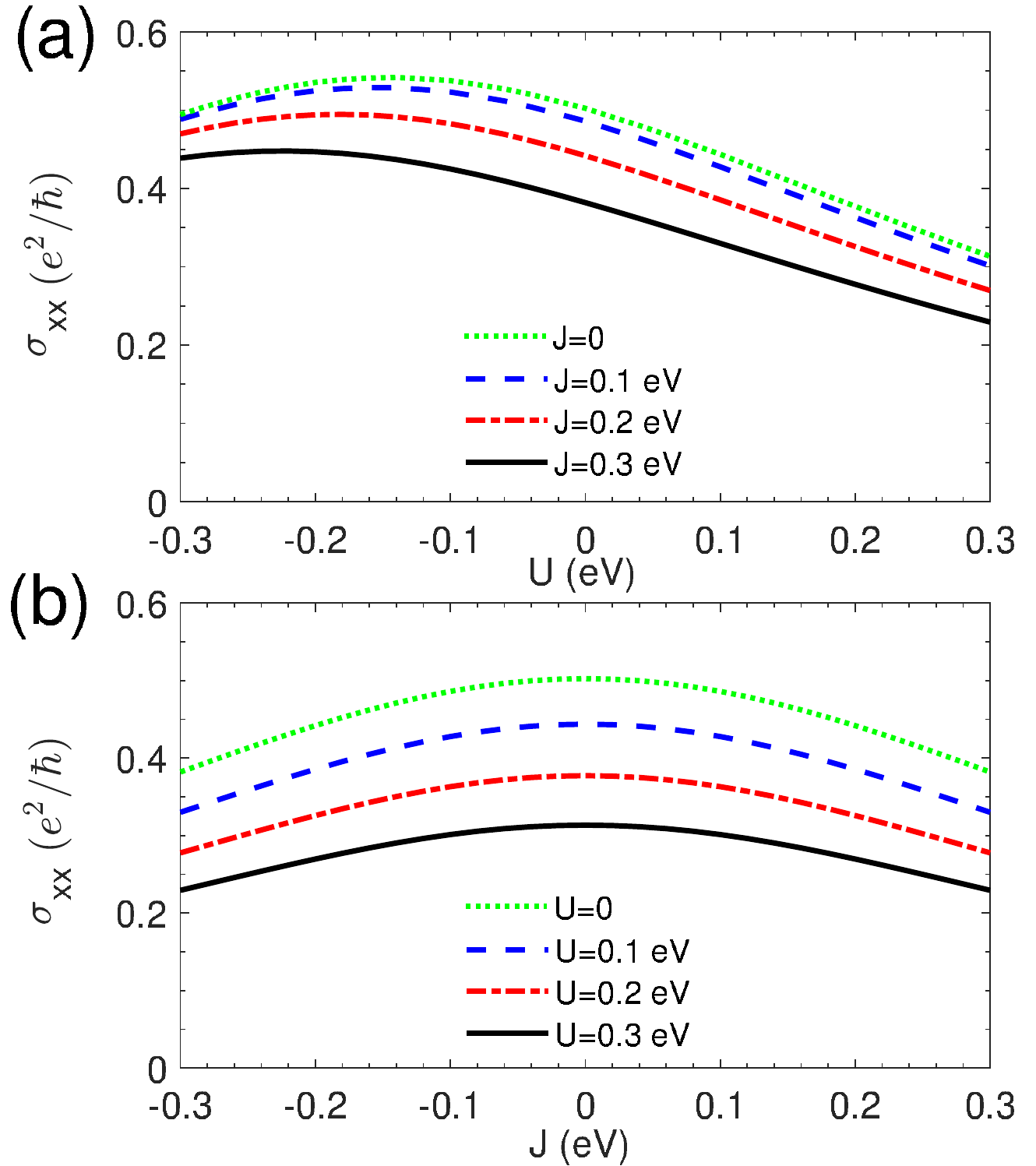}}
\caption{Calculated conductivities as a function of (a) nonmagnetic potential $U$ and (b) magnetic potential with magnetization ${\bf m}$ along $z$-axis. The other parameters are $\Delta=0$, $\omega=0.05$\,eV, $\lambda=250$\,eV$\cdot$\AA$^3$, and  $n_{imp}=0.05$.}
\label{F6}
\end{figure}
\begin{figure}
\centerline{\includegraphics[scale=0.6]{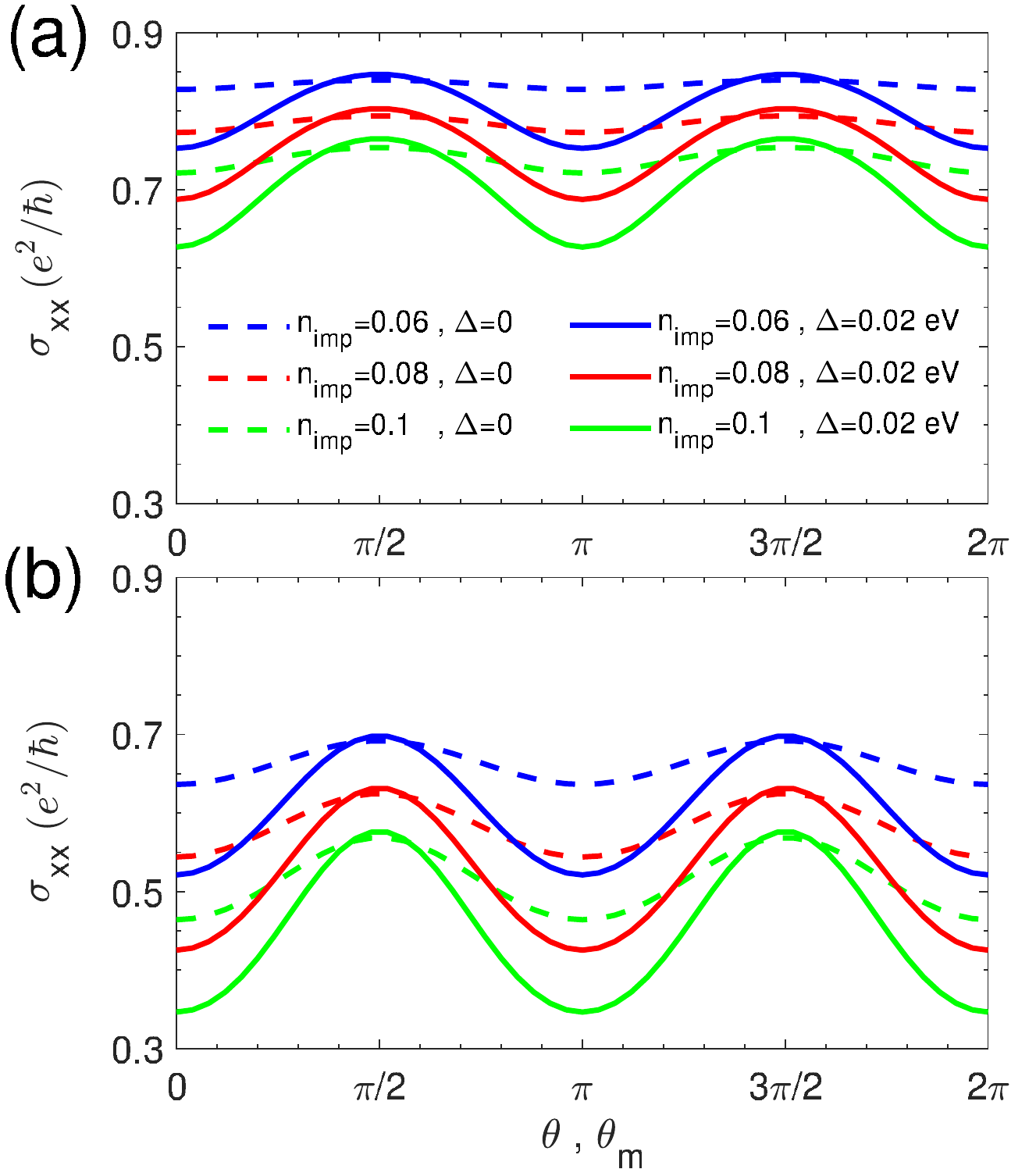}}
\caption{Calculated conductivities as a function of magnetization angle $\theta_{m}$ and $\theta$ at impurity potentials (a) $U=0$ and (b) $U=0.2$\,eV. The other parameters are $J=0.2$\,eV , $\omega=0.1$\,eV, $\lambda=250$\,eV$\cdot$\AA$^3$.}
\label{F7}
\end{figure}
Now we consider the electric dc conductivity of Dirac electrons along the longitudinal direction on the surface of topological insulators, doped with magnetic and/or chemical (nonmagnetic) dopants. In Figs. 4(a) and 4(b) we have depicted the conductivity $\sigma_{xx}$ of doped surface as a function of energy in the absence and presence of warping effect, respectively.  For the clean surface, $\sigma_{xx}(\omega)$ is also shown for comparison in the inset of Fig. 4(b). In the case of $\lambda=0$, the conductivity of clean surface exhibits a V-shaped behavior, analogous to the surface DOS, reflecting the linear dispersion of Dirac electrons (see Figs. 2(a) and 2(b) for $n_{imp}=0$). In contrast, the inclusion of warping effect in the Hamiltonian deviates the conductivity of clean surface at $\omega\geq$ 0.15 eV from linearity to parabola-like behavior, causing an increase in $\sigma_{xx}(\omega)$ compared to the conductivity of surface with $\lambda=0$. Such a deviation can be understood by considering the relation between conductivity and velocity operator as $\sigma_{xx}\propto |v_{x}|^{2}$ (see Eq. (\ref{eq19})) with $v_{x}=v_{F}\sigma_{y}+3\lambda k^{2}\cos(2\phi)\,\sigma_{z}$. It is clear that $v_{x}$ increases by including warping effect in our model. Although the influence of $\lambda$ in velocity increase at low energies is negligible, the increase in energy which manifests itself in the $k$ parameter of velocity equation increases the conductivity compared to the case of $\lambda=0$. Moreover, as one can see from Fig. 1(a), by increasing the electron energy, the absolute value of energy slope with respect to $k_{x}$ along the CEC enhances, causing the increase in $\sigma_{xx}(\omega)$ compared to that in the case of $\lambda=0$.

On the other hand, Fig. 4(a) shows that for the case of $\lambda=0$, the electric conductivity strongly depends on the impurity potentials $J$ and $U$. In fact, when the impurities are introduced, $\sigma_{xx}(\omega)$ exhibits a non-monotonic behavior with respect to energy as a result of induced resonant states. When the impurities are solely magnetic, the conductivity demonstrates a symmetrical behavior in terms of $\omega$, whereas the conductivity becomes asymmetric when nonmagnetic impurities are introduced. The impurity potential $U$ shifts the minimal conductivity to the higher-energy region, similar to the feature seen in DOS spectrum, shown in Fig. 3. In the presence of warping effect, however, $\sigma_{xx}(\omega)$ enhances monotonically as $|\omega|$ is increased from the Dirac point (see Fig. 4(b)). As mentioned above, such a conductance increase is due to the warped CEC, causing an enhancement in electron group velocity. For instance, in the presence of warping effect, the conductivity of surface with impurity potentials $J=0.6$ eV and $U$=0 reaches $\sim 1.1e^{2}/\hbar$ at $|\omega|=0.3$eV, while in the absence of warping effect, $\sigma_{xx}$ is $\sim 0.2e^{2}/\hbar$ at the same energy. We note that the introduction of magnetic impurities with ${\bf{m}}$ in the $z$-direction opens up a band gap in $\sigma_{xx}(\omega)$ at the Dirac point as a result of time-reversal symmetry breaking, regardless of $\lambda$ strength.

The electric conductivity can be affected by magnetization orientations ${\bf{m}}$ and ${\bf{M}}$. To see this, in Figs. 5(a) and 5(b), we have shown $\sigma_{xx}$ versus magnetic impurity potential $J$, when magnetizations are aligned in the $z$-  and $x$-directions.  The increase in the impurity potential $J$ and/or the impurity concentration $n_{imp}$ suppress the dc conductivity gradually, due to the increase of electron scattering from the impurity atoms, which decreases the electrical conductivity. Moreover, by applying an external magnetic field along the $z$-direction, $\sigma_{xx}$ decreases as a result of band gap opening in the energy dispersion (compare the blue lines in Fig. 5(a) and 5(b)). It is worth mentioning that although the magnetizations along $x$-direction do not break time-reversal symmetry, they can shift the Dirac cone in momentum space, accompanied by a deformation of CEC \cite{Arabikhah2019}.

The competition between impurity potentials in generating higher conductivity determines the optimal parameters of $U$ and $J$. In Figs. 6(a) and 6(b), we have plotted the surface conductivity versus nonmagnetic and magnetic impurity potentials at different $J$ and $U$ values, respectively. In the absence of magnetic impurities, $\sigma_{xx}$ exhibits the highest values at the given $U$ interval with a maximum of 0.54 $e^{2}/\hbar$ at $U\simeq -0.15$\,eV. As $J$ is increased, the conductivity decreases along with a shift in its maximum value toward more negative $U$ values (see Fig. 6(a)). This occurs because $\sigma_{xx}$ is a function of $\omega$ and that the energy-gap opening depends on $J$ value. In contrast, the conductivity versus $J$ is symmetrical because the second term in Eq. (\ref{eq3}) is proportional to $\sigma_{z}$ and that the increase in $|J|$ opens up a gap at the Dirac point, as shown in Fig. 4. We can see in Fig. 6(b) that for all $U$ values, $\sigma_{xx}(\omega=0.05$\,eV) versus $J$ is maximum at $J=0$, due to the time-reversal symmetry. The increase in the impurity potential $U$ reduces the conductivity as a result of shifting the Dirac point to the positive energy region.

Note that in this study to explore the surface electronic states and conductivity of doped topological insulators, we have considered various values of the potential strengths $U$ and $J$. Since the magnetic and chemical impurities are independent, $J$ can be stronger or weaker than $U$ values, depending on the ion species occupying the position ${\bf r}_n$ on the surface of topological insulator (see Fig. 1(b)). As mentioned earlier, the exchange coupling $J$ is proportional to the spin $S$ of magnetic impurities. Therefore, magnetic impurities with $d$ or $f$ electrons can give rise to a large total spin quantum number and even larger than chemical potential strength $U$ for some nonmagnetic impurity ions.    

The change in magnetization orientations can also exhibit a significant influence on $\sigma_{xx}$ values. To show this we have depicted in Figs. 7(a) and 7(b) the conductivity as a function of magnetization angles in the absence and presence of impurity potential $U$, respectively.
The electric conductivity demonstrates an oscillatory behavior in terms of $\theta$ and $\theta_{m}$, as a result of gap opening and gap closing when ${\bf M}$ and ${\bf m}$ are rotated continuously from $z$ to $x$ axis in the $x$-$z$ plane.  Although, the increase of impurity concentration which enhances the electron scattering reduces the conductivity, the amplitude of oscillations remains almost unchanged. The oscillation of $\sigma_{xx}$ values is very small in the absence of magnetic proximity effect. However, as shown by solid lines in Fig. 7, the amplitudes of oscillations are enhanced not only by a magnetic proximity effect but also by nonmagnetic impurities, indicating that the conductivity on the surface of topological insulators can be manipulated by tuning the directions of external magnetization and impurity magnetic moments.

It is worth mentioning that the present model can be applied to the surface of topological insulators where a thin magnetic layer is created at their surface by a low-energy Cr ion beam \cite{Cortie2020}. The robust gapless surface states against surface magnetic impurities with controlled densities in thin films of some topological insulators, such as (Bi$_{0.5}$Sb$_{0.5})$Te$_{3}$, can also be studied using the present theory \cite{Yu2020}.

\section{Conclusion}
Using a self-consistent $t$-matrix approximation, we have investigated the electronic surface states and dc conductivity of Dirac electrons in topological insulators with warping effect. Our findings show that when the warping effect is absent, an increase in impurity concentration leads to an increase in the resonant peaks, suggesting that the electronic states in the vicinity of the Dirac point can be tuned by varying the density of magnetic impurities.
When the warping term is introduced, the resonant states induced by impurities almost disappear, regardless of the dopant type. The particle-hole symmetry at low energies is always preserved in topological insulators with strong enough warping effect, such as Bi$_{2}$Te$_{3}$. We found that the warped surface states enhance the electric conductivity of Dirac electrons in both doped and clean surfaces, due to the enhancement of electron group velocity. We show that the surface conductivity of topological insulators can be controlled by tuning the doping type, impurity concentration, the magnetization direction of applied field and that of the magnetic dopants.

\end{document}